# Orientations of linear stone arrangements in New South Wales

Duane W. Hamacher[1], Robert S. Fuller[2] and Ray P. Norris[2,3]

[1]Nura Gili, University of New South Wales, Sydney, NSW 2052, Australia
[2]Department of Indigenous Studies, Macquarie University, NSW 2109, Australia
[3]CSIRO Astronomy and Space Science, PO Box 76, Epping, NSW 1710, Australia
Corresponding email: d.hamacher@unsw.edu.au

## Abstract

We test the hypothesis that Aboriginal linear stone arrangements in New South Wales (NSW) are oriented to cardinal directions. We accomplish this by measuring the azimuths of stone arrangements described in site cards from the NSW Aboriginal Heritage Information Management System. We then survey a subset of these sites to test the accuracy of information recorded on the site cards. We find a preference recorded in the site cards for cardinal orientations among azimuths. The field surveys show that the site cards are reasonably accurate, but the surveyors probably did not correct for magnetic declinations. Using Monte Carlo statistics, we show that these preferred orientations did not occur by chance and that Aboriginal people deliberately aligned these arrangements to the approximate cardinal directions. We briefly explore possible reasons for these preferred orientations and highlight the need for future work.

## Introduction

This paper represents the first rigorous study of the orientations of stone arrangements in New South Wales (NSW), focusing on linear arrangements ('stone rows'). Specifically, we test the hypothesis that linear stone arrangements have a preferred orientation to cardinal directions. We accessed site cards from the NSW Aboriginal Heritage Information Management System (AHIMS) database and filtered them through a rigorous selection process to reject those with insufficient information from which to determine stone row orientations. We then measured the orientations of each stone row described in the site cards. To test the accuracy of information recorded on site cards, a subset of sites were also revisited and surveyed. Monte Carlo statistics were used to test whether or not any preferred orientation amongst linear stone arrangements is the result of chance. Finally, we discuss future work and explore the causes for any preferred orientations.

## Cardinal Directions in Aboriginal Languages

The concept of cardinal directions is found among several of the hundreds of Aboriginal language groups in Australia (e.g. Breen 1992, 1993; Edmonds-Wathen 2011; Haviland 1996; Kirton 1987; Laughren 1978, 1992; McGregor 1990; Yallop 1977). While many Aboriginal languages contain names or concepts for four cardinal directions, some languages contain as many as five or six (Laughren 1978; Nash 1980; Spencer and Gillen 1899). Of particular interest with respect to Aboriginal astronomy is whether these directions are based on an abstract concept of relative space, or an absolute concept based on features of the landscape (e.g. Lewis 1976), wind directions, river flow directions or the rising/setting position of the sun.

Guugu Yimithirr speakers from Queensland (Qld) describe the relative position of objects or places in terms of root words that represent the cardinal directions in four general quadrants (Haviland 1979, 1993, 1996): north (*gungga*), east (*naga*), south (*jiba*) and west (*guwa*). In terms of gauging directions by solar positions, Kunwinjku speakers of the Northern Territory (NT) may refer to absolute directions in terms of sunrise and sunset, corresponding to *abalkbang manyij* (east) and *wurrying manyij* (west), respectively (Edmonds-Wathen 2011:222). Several language groups in Central Australia have particular words for the cardinal directions, most notable of which are the Warlpiri, who have an entire culture based on a system of cardinal directions (Laughren 1978, 1992; Nash 1980). Interestingly, Breen (1993) found that an Alyawarr community in the NT that had migrated to a different region altered the names of the cardinal points by 90°. The reason for this is currently unknown, but Breen noted that the Wangkumara (Qld) terms for east and west are based on *mirla*, meaning sun. The linguistic relationship between east/west and the sun is also found in the Yirandhali and other Mari languages of Qld (Tindale 1938/39). In the Yirandhali language, the term for east is *kunggari*, meaning literally 'sun get up' (Breen 1993; Tindale 1938/39). An identical concept is found near Lake Boga, Victoria (Vic.), where east is *worwalling gnowie*, meaning 'where the sun rises' and west is *purticalling gnowie*, meaning 'where the sun sets' (Stone 1911:451).

From this, it is clear that east and west are denoted by the rising and setting sun. However, it is not clear how exact these cardinal directions are to scientific definitions of cardinal directions based on the rotational axis of the earth. Breen (1993) tested the accuracy of various words representing cardinal directions in several different areas by asking Aboriginal people and found that the direction, as compared to a compass, ranged rather dramatically—up to 90° in some cases, leading Breen to suggest that the names for cardinal directions represent a vague area rather than an exact direction, and it would be unreasonable to expect a person to 'even try to be exact' (Breen 1993:28). Breen also concluded that, if people were placed in a new area with an unfamiliar topography, they would require the sun to determine direction by during the day, or the stars at night. If dropped in an unfamiliar area on a cloudy day, one would essentially be unable to determine the cardinal points (Breen 1993:27). In some regions, cardinal directions play a role in burial rituals. In NSW, for example, graves were found with the dead buried in a sitting position, facing east (Dunbar 1943; Mathews 1904:274).

**Stone Arrangements**

Common to many Aboriginal cultures were stone arrangements of various designs and morphologies, including circles, lines, pathways, standing stones and cairns, with purposes that ranged from practical (e.g. fish traps, land boundaries) to mythological, and ceremonial/ritual (e.g. initiation or burial; Flood 1999a).

Stone arrangements vary in size from a few meters to hundreds or even thousands of meters in length or diameter and are typically constructed from local rocks that are small and movable by one or two people, although occasionally they can weigh as much as 500 kg (Lane and Fullagar 1980; Long and Schell 1999). The ages of stone arrangements are unknown, but smaller arrangements are likely on the order of hundreds of years old, as sedimentation and disruption by natural processes would likely have buried or destroyed older arrangements.

McCarthy (1940) suggested that stone arrangements used for ceremonial purposes incorporate the surrounding landscape, and may indicate the direction of a landmark or mimic a land feature. The results of the current study indicate that many of the stone arrangements examined were on a hill or a location of higher elevation that commanded a panoramic (full or partial) view of the surrounding landscape. Flood (1999b:239-240) notes that in NSW, elevated sites were preferred for ceremonies, such as male initiation, or Bora, ceremonies. Bora sites were generally made of stone or raised earth in the form of two circles connected by a pathway (a project measuring the orientations of Bora sites in southeastern Australia is underway by the authors).

Some researchers have noted that stone arrangements align to cardinal points in southeastern Australia and Tasmania. Examples include Black (1945:212-213) and Flood (1999b:208) who both describe linear stone arrangements oriented to a North/South direction and Bartholomai & Breeden (1961:233) and Winterbotham (1957:38) who describe ceremonial stone arrangements oriented to the East/West or North/South. Other examples include the Wurdi Youang and Carisbrook stone arrangements in Vic., which are oriented to the cardinal points (Hamacher and Norris 2011; Massola 1963).

For a further treatise on stone arrangements, the interested reader is referred to Attenbrow (2002), Bannerman and Jones (1999), Black (1950), Campbell and Hossfield (1964), Dow (1939), Enright (1937), Frankel (1982), Kelly (1968), Lane (2009), Lane and Fullager (1980), Long and Schell (1999), Love (1946), Massola (1968b), McCarthy (1940), Palmer (1977), Ross (2008), Ross & Ulm (2009), Rowlands & Rowlands (1970), Towle (1939a, 1993b) and Winterbotham (1949).

## Methods

### Archaeological Site Cards

Aboriginal sites and artefacts in NSW are registered with AHIMS and administered by the NSW Office of Environment and Heritage. Information about each site is stored on archaeological site cards submitted by both professionals and amateurs: these can vary significantly in the quality and quantity of information provided. Therefore, the accuracy of the data collected is difficult to assess without physically resurveying each site. Not all existing stone arrangements are registered with AHIMS and some sites cards are restricted owing to their cultural sensitivity; no restricted sites were included in this study.

We examined 643 stone arrangement site cards from AHIMS, developing a rigorous selection process to cull sites that did not provide useful data. We rejected sites that did not meet all of the following five criteria:

1. The arrangement must be clearly Aboriginal in origin[1];

---

[1] Rows of stone were sometimes constructed as survey markers or borders by Europeans: these are generally called survey lockspits. These can mimic Aboriginal stone rows and some have been identified along the NSW/ACT

2.  The stones must form an unambiguous linear pattern;
3.  The arrangement must consist of at least five stones of comparable size;
4.  The linear pattern of stones must not clearly be a smaller component of a larger, non-linear structure, such as a circle, animal motif or other pattern; and,
5.  A map or directions and a photo, diagram or a sufficient description of the site must be available.

Of the 643 AHIMS site cards examined, 618 were rejected as not meeting the criteria, leaving 24 sites consisting of 32 stone rows (Table 1); parallel rows ('pathways') were categorized as a single row. Five of the sites contained multiple stone rows (12-2-0028, 12-5-0056, 28-2-0005, 44-6-0019, 45-4-0217). Although site card 44-6-0019 shows a stone circle, the three stone rows within the circle (two parallel rows to the north and one to the east) were included because they were distinct within the arrangement and not merely a straight component of the overall circle. Locations of the sites used in this study are given in Figure 1 and sketches of the stone rows, as shown in the site cards, are given in Figure 2. Some site cards provided orientations but not sketches.

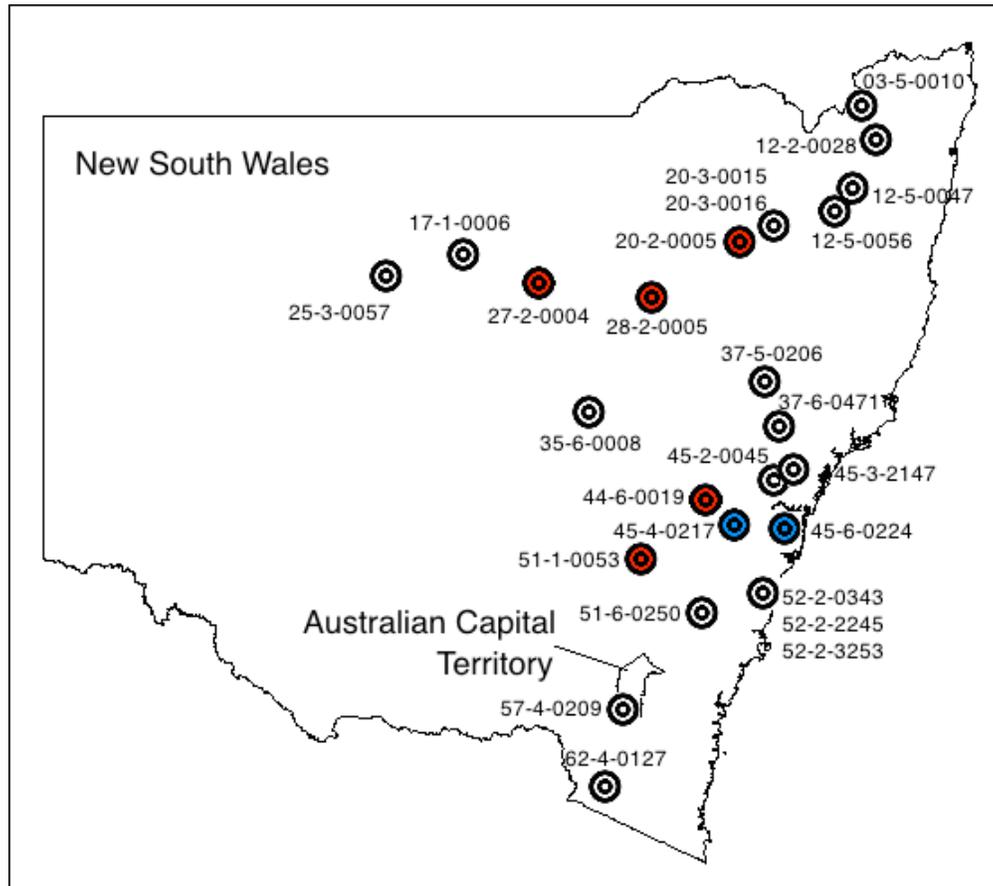

**Figure 1:** Locations of linear stone arrangements selected for this study, including those that were surveyed (red) and those that were visited but could not be located or were previously destroyed (blue).

border. All stone rows in our final dataset were crosschecked against the locations of known lockspits and rejected if they were discovered to be one.

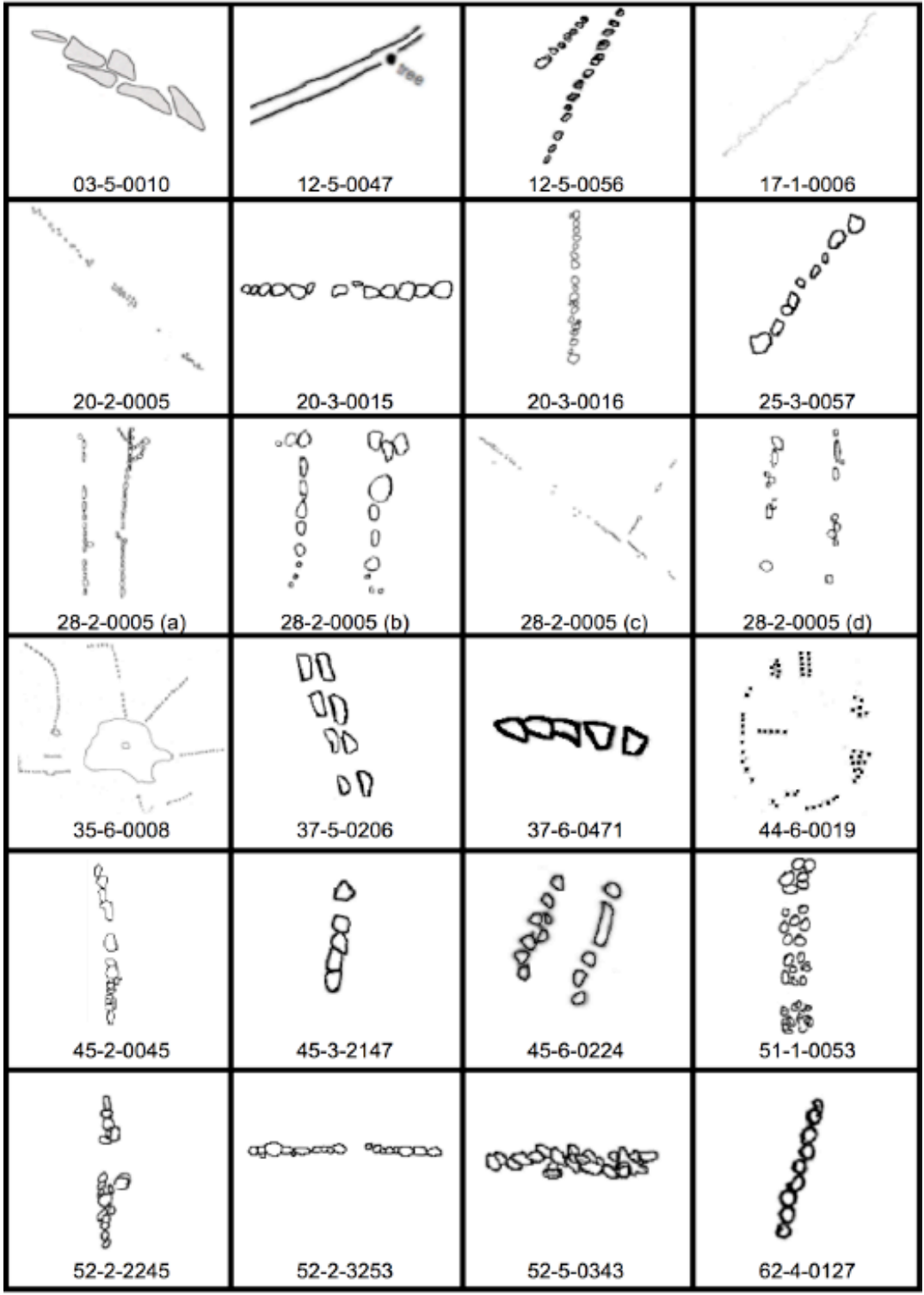

**Figure 2:** Stone arrangement sketches taken from site cards. Stone rows have been cropped from the rest of the sketch in most cases (except 44-6-0019). Site 45-2-0045 was removed when it proved to be a lockspit and 35-6-0008 was rejected upon visiting the site and finding that the site card orientation was incorrect and vegetation overgrowth concealed the arrangements. Sites that provided only a description or a photo were not included. North (as labelled in the site card) is to the top of the page.

**Table 1:** The general locations of the 24 sites in this study. Coordinates provided are for a location within 10 km of the site so as to protect their specific location but give the reader an overall impression of their distribution. in the site card) is to the top of the page.

| Site Card | Zone | Latitude | Longitude | Elevation | N |
|---|---|---|---|---|---|
| | | (°) | (°) | (m) | |
| 03-5-0010 | 56 | −28.8 | 152.1 | 960 | 1 |
| 12-2-0028 | 56 | −29.1 | 152.3 | 980 | 2 |
| 12-5-0047 | 56 | −29.6 | 152.2 | 880 | 1 |
| 12-5-0056 | 56 | −29.8 | 152.1 | 1100 | 2 |
| 17-1-0006 | 55 | −30.9 | 146.6 | 200 | 1 |
| 20-2-0005 | 56 | −30.4 | 150.9 | 820 | 1 |
| 20-3-0015 | 56 | −30.2 | 151.3 | 1000 | 1 |
| 20-3-0016 | 56 | −30.2 | 151.3 | 1000 | 1 |
| 25-3-0057 | 55 | −31.2 | 145.4 | 240 | 1 |
| 27-2-0004 | 55 | −31.3 | 147.6 | 160 | 1 |
| 28-2-0005 | 55 | −31.3 | 149.3 | 540 | 5 |
| 37-5-0206 | 56 | −32.6 | 150.9 | 480 | 1 |
| 37-6-0471 | 56 | −32.9 | 151.0 | 160 | 1 |
| 44-6-0019 | 55 | −33.5 | 149.9 | 820 | 1 |
| 45-3-2147 | 56 | −33.2 | 151.1 | 180 | 1 |
| 45-4-0217 | 56 | −33.7 | 150.5 | 460 | 2 |
| 45-6-0224 | 56 | −33.7 | 151.2 | 420 | 1 |
| 51-1-0053 | 55 | −34.0 | 149.0 | 700 | 1 |
| 51-6-0250 | 55 | −34.7 | 150.0 | 680 | 1 |
| 52-2-2245 | 56 | −34.5 | 150.7 | 500 | 1 |
| 52-2-3253 | 56 | −34.5 | 150.7 | 460 | 1 |
| 52-5-0343 | 56 | −34.5 | 150.7 | 560 | 1 |
| 57-4-0209 | 55 | −35.8 | 148.9 | 1660 | 1 |
| 62-4-0127 | 55 | −36.7 | 148.5 | 300 | 1 |

### *Site Card Measurements*

Using information provided on the site cards, we measured the azimuth of each stone arrangement (Table 2) using the compass azimuth if provided or by using a protractor and ruler on graph paper. Azimuths are given between 0º and 179º, with 0º/180º representing north/south, and 90º representing east/west. Azimuths measured with a magnetic compass (i.e. that measures direction with respect to the magnetic north pole) in NSW are between 9º and 13º greater than their azimuth with respect to true (geographic) north. The difference between true and magnetic north is called 'magnetic declination' (Dmag). We used an online program provided by Geoscience Australia (2011) to calculate Dmag using the geographic coordinates provided in the site cards, the elevation taken from Google terrain maps (with a relief of 20 m) and the date of the original site recording as noted on the site card. Dmag was calculated for every site and then subtracted from the measured azimuth, (Table 2); however, only five of the 24 site cards discriminate between true and magnetic north (the latter by indicating 'MN'). The remaining site cards either labeled north as 'N' or stated the orientation with no reference as to whether magnetic declination had been corrected for. For this reason, we only use the corrected azimuths (i.e. those determined by subtracting Dmag from the azimuth) for five sites in the analysis.

**Table 2:** Linear stone arrangements remaining after the application of selection criteria. The five site cards with an asterix denoted north as magnetic.

| Site Card | $Az_{SC}$ (°) | $D_{mag}$ (°) | $Az_C$ (°) | Date | Site Card | $Az_{SC}$ (°) | $D_{mag}$ (°) | $Az_C$ (°) | Date |
|---|---|---|---|---|---|---|---|---|---|
| 03-5-0010* | 120 | 11.112 | 109 | Jun-77 | 37-5-0206 | 157 | 11.958 | 145 | Apr-94 |
| 12-2-0028 | 0 | 11.277 | 169 | Oct-97 | 37-6-0471 | 101 | 12.101 | 89 | Oct-90 |
|  | 90 | 11.277 | 79 |  | 44-6-0019 | 0 | 11.970 | 168 | Sep-78 |
| 12-5-0047* | 65 | 11.388 | 54 | Nov-95 |  | 90 | 11.970 | 78 | Sep-78 |
| 12-5-0056 | 27 | 11.428 | 16 | Sep-97 | 45-3-2147 | 2 | 12.242 | 170 | Jun-93 |
|  | 43 | 11.428 | 32 |  | 45-4-0217 | 0 | 12.238 | 168 | Nov-92 |
| 17-1-0006 | 47 | 10.071 | 37 | Jan-91 |  | 90 | 12.238 | 78 |  |
| 20-2-0005 | 134 | 11.178 | 123 | May-81 | 45-6-0224 | 16 | 12.467 | 4 | Apr-92 |
| 20-3-0015* | 178 | 11.301 | 167 | Aug-79 | 51-1-0053 | 0 | 11.801 | 168 | Dec-93 |
| 20-3-0016* | 90 | 11.301 | 79 | Aug-79 | 51-6-0250 | 90 | 12.393 | 78 | Jul-04 |
| 25-3-0057 | 38 | 9.738 | 28 | Jun-79 | 52-2-2245 | 0 | 12.552 | 167 | Jul-98 |
| 27-2-0004* | 90 | 10.496 | 80 | Feb-91 | 52-2-3253 | 90 | 12.552 | 77 | Jul-98 |
| 28-2-0005 | 1 | 11.005 | 170 | Dec-79 | 52-5-0343 | 90 | 12.552 | 77 | Jul-98 |
|  | 3 | 11.005 | 172 |  | 57-4-0209 | 0 | 12.415 | 168 | May-06 |
|  | 0 | 11.005 | 169 |  | 62-4-0127 | 15 | 12.616 | 2 | Jul-81 |
|  | 32 | 11.005 | 21 |  |  |  |  |  |  |
|  | 125 | 11.005 | 114 |  |  |  |  |  |  |

*Field Survey Methods*

To test the accuracy of information recorded on the site cards, we surveyed a subset of the 24 sites. Of these, we were unable to gain access to 17, either because of their remote location or our inability to contact or obtain permission from the traditional owners, the current landowners or NSW Parks and Wildlife. Sites 12-2-0028 and 45-6-0224 had been previously destroyed, Site 44-6-0019 was damaged, we were unable to physically relocate Site 45-4-0217 (which may have been destroyed since initial recording) and dense vegetation had grown over Site 35-6-0008 such that it was impossible to clearly identify the stones in any of the several arrangements noted on the site card (accordingly this site was also excluded from analysis). The sketch for Site 35-6-0008 did not indicate north, but it was assumed to be at the top of the page. However, upon visiting the site this assumption was proven incorrect based on the orientation of a large rock in the centre of the site. As such, Site 35-6-0008 was also excluded from analysis. Given time constraints, we were unable to survey three of the five rows at Site 28-2-0005 in detail. We surveyed five sites (Table 3) using a standard military lensatic compass, a Sokkisha C40 (D10355) automatic level (dumpy), stadia rod, tape measures and a hand-held GPS as follows:

1. The stones on each end of the linear arrangement were identified.
2. The leveled dumpy was placed 1 m from one end-stone along the axis of the end-stones, using a weighted string over the point on the ground to ensure accuracy.
3. The stadia rod was placed on the centre of the opposite end-stone.
4. Using the sight on the dumpy as an alidade, the magnetic azimuth of the stadia rod was measured by aligning the compass sight-wire along the dumpy sight to the edge of the staff.
5. Using the dumpy and stadia rod, the topographic relief was recorded along the stone row.
6. Each stone was numbered and its dimensions recorded.
7. The position of the centre of each stone was recorded relative to the line connecting the end-stones.
8. The site was photographed.
9. The presence of Aboriginal artefacts or art in the area was noted.

For the purposes of analysis the azimuths of parallel stone rows were averaged. This data is provided in Table 2, where all measured azimuths are rounded to the nearest degree (errors are discussed in the following section).

### Error Analysis

The two general errors that affect the measurements of stone rows are (1) site card errors and (2) our field survey errors. Site card errors include the accuracy of the original sketch, the accuracy of the survey methods and the skill of the surveyor. These errors are difficult to quantify without visiting the site. Some site card drawings are not to scale and represent only rough sketches. However, if the sketch of the site is taken to be accurate, the main source of error stems from physically measuring the azimuth given in the site card using a ruler and protractor. Both the site cards and our field surveys are subject to the same three general errors in measuring the azimuth, as described below.

#### Errors in Measuring the Magnetic Azimuth

The azimuth of the stone row surveyed in the field was measured by taking several compass bearings and averaging them. The compass bearings varied by a maximum of ~2º, which we consider to be the 'human error estimate'.

**Table 3:** A list of the sites. Sites surveyed = X; sites visited but not surveyed = V.

| Site Card | Accessibility | Cardinal | Surveyed | Condition | MN |
|-----------|---------------|----------|----------|-----------|----|
| 03-5-0010 | None | | | Unknown | X |
| 12-2-0028 | None | X | | Destroyed? | |
| 12-5-0047 | None | | | Unknown | X |
| 12-5-0056 | None | | | Unknown | |
| 17-1-0006 | None | | | Unknown | |
| 20-2-0005 | Granted | | X | Good | |
| 20-3-0015 | None | X | | Unknown | X |
| 20-3-0016 | None | X | | Unknown | X |
| 25-3-0057 | None | | | Unknown | |
| 27-2-0004 | Granted | X | X | Good | X |
| 28-2-0005 | Granted | X | X | Good | |
| 37-5-0206 | None | | | Unknown | |
| 37-6-0471 | None | | | Unknown | |
| 44-6-0019 | Granted | X | X | Damaged | |
| 45-3-2147 | None | X | | Unknown | |
| 45-4-0217 | Granted | X | V | Unidentified | |
| 45-6-0224 | Granted | X | V | Destroyed | |
| 51-1-0053 | Granted | X | X | Good | |
| 51-6-0250 | None | X | | Unknown | |
| 52-2-2245 | None | X | | Unknown | |
| 52-2-3253 | None | X | | Unknown | |
| 52-5-0343 | None | X | | Unknown | |
| 57-4-0209 | None | X | | Unknown | |
| 62-4-0127 | None | | | Unknown | |

*Errors in Magnetic Declination*

The Earth's magnetic field varies over time, causing the Dmag to shift at any given location by ~0.3° over a 25 year period. The azimuths recorded in the site cards were recorded as early as 1977 (33 years prior to our resurvey), thus giving an azimuthal error of 0.3°. If the location and elevation of the site and the date of the survey are known, the azimuthal error is reduced to only ~0.001°. However, local magnetic anomalies can cause variations of up to 4° (Goulet 2001), such as the influence of iron-bearing stone.

The Great Dividing Range of eastern Australia, where many of the stone arrangements are found, is volcanic in origin. Volcanic (igneous) rocks, such as basalt, are typically iron-rich and in fact, many of the stones forming the arrangements are composed of basalt. Therefore, we use an upper error estimate of 4° to account for this effect.

*Errors in Measuring the Dimensions of Each Stone*

The dimensions of each stone were recorded, including their maximum width, length and height (with respect to the ground surface). The width and length are relative to the x- and y-axes, as determined by the orientation of the two end-stones. The positions of each stone were recorded with respect to the midpoint of the width and length. Since measurements are taken from the centre of each stone with an accuracy of ±0.02 m, the azimuthal error for an arrangement with an average length of 7.5 m is ~0.3°.

*Final Error Estimates*

The human error (2°), the error from local magnetic anomalies (4°), the error in magnetic declination (0.001°) and the error in measuring the dimensions of the stones (0.3°) are used to calculate an error of ~5° (±2.5°) for the surveyed azimuth ($\sigma_{mag}$). To account for this error in the analysis, azimuths are grouped into 5° bins. Because the value of $\sigma_{mag}$ is dominated by magnetic anomalies and human error, we apply an error of 5° to orientations of stone arrangements in the site cards. However, we do not know how accurate the surveyors were in recording the azimuths, so this can only be considered an approximate estimate based on our quantified estimates.

**Results and Statistical Analysis**

*Site Card Measurements*

The site card data consisted of 32 azimuths from 24 sites, mapped as a histogram shown divided into 5° bins centered at 0° (N/S) and 90° (E/W), as shown in Figure 3. As demonstrated, there is an unambiguous preference for cardinal orientations, with nine azimuths in the N/S bin and seven in the E/W bin. To test whether or not these peaks could occur by chance, we use Monte Carlo statistics. Monte Carlo statistics sample probability distributions to produce millions of possible outcomes, which is useful for determining the likelihood of something occurring by chance (Fishman 1995; Vose 2000).

A Monte Carlo simulation was performed in which 39 random angles were assigned to any one of the 36 5° bins from 0° to 180°. After running one billion ($10^9$) simulations, in only ~6000 of the simulations did any bin contain nine or more azimuths, showing that the probability of getting a peak of nine in any one bin is approximately $6\times10^{-6}$.

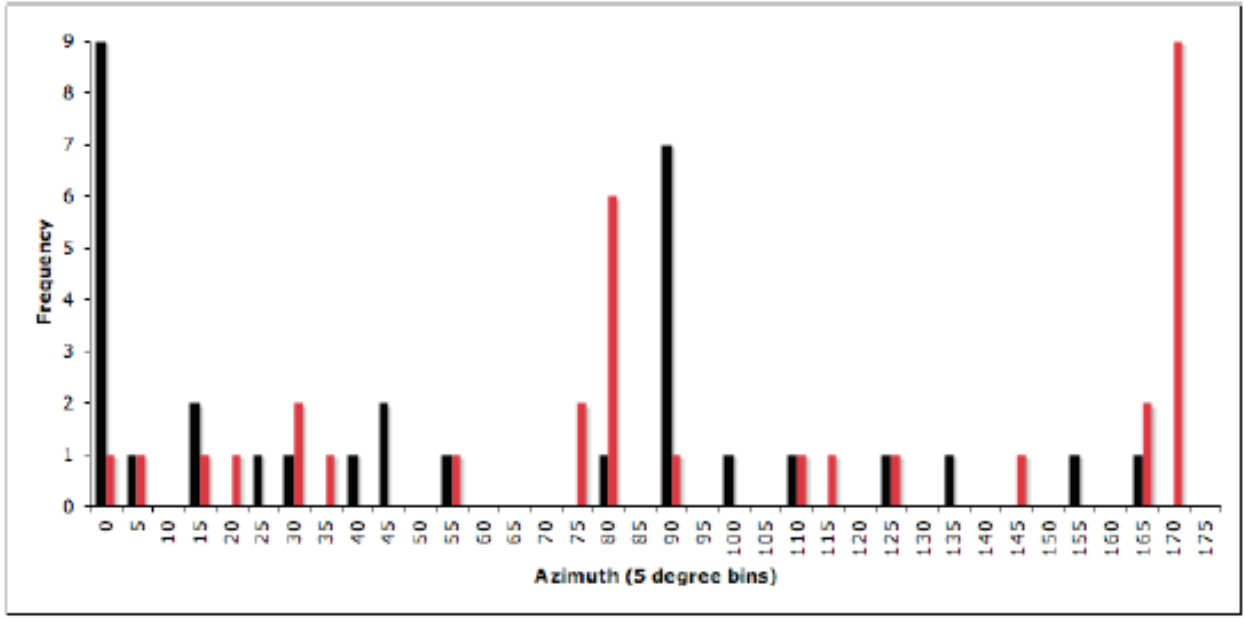

**Figure 3:** Histogram showing the azimuths of stone rows from the site cards. Black: Only site cards that cite north as magnetic north are corrected for Dmag. Red: All site cards are assumed to show magnetic north, so all azimuths are corrected for Dmag. Measurements have an estimated error of 5°, which is why they are given in 5° bins.

The site card azimuths did not attain a peak of nine in any random bin, but in the bin centered at 90°. This occurred in only 1600 of the one billion simulations, implying that the probability of a peak centered at 90° occurring by chance is $1.6 \times 10^{-6}$. In none of the one billion simulations were peaks of nine or more obtained in both the 0° and 90° bins. We estimate the probability of obtaining two such peaks (at 0° and 90°) to be approximately $2 \times 10^{-12}$. From this, we conclude that if the site cards are accurate, the preferred orientations of the stone rows are not the result of chance alignments.

*Field Survey Measurements*

To check the accuracy of information reported on the site cards, we surveyed a subset of six stone arrangements at five sites. Table 4 gives the azimuths for the six arrangements. To compare the accuracy of information on the site cards with the resurveyed data, we calculated the mean ($\theta$) and standard deviation ($\sigma_{Az}$) of the difference between the surveyed azimuth and the site card azimuth ($\Delta Az = Az_{SC} - Az_F$). Given six data points for $\Delta Az$ from Table 4, $\theta = 19.5°$ and $\sigma_{Az} = 11.5°$.

The magnetic declination at the sites resurveyed ranged from 10.5° to 12.0°, with a mean ($\mathbf{D}_{mag}$) of 11.3° and a standard deviation ($\sigma_{Dmag}$) of 0.54°. Because $\mathbf{D}_{mag}$ is within 1-$\sigma$ of $\theta$, it seems the site cards were probably not generally corrected for $D_{mag}$ by the surveyors (i.e. most of the surveyors probably recorded the magnetic compass bearing rather than true North). The surveyed azimuths represent a small percentage of the site card azimuths (~15%), so the remaining sites need to be surveyed to determine more accurately whether the arrangements are aligned to true or magnetic north.

**Table 4:** Data for all surveyed linear stone arrangements surveyed. m = the slope of the linear best fit, $\chi^2$ is the chi-squared distribution, $\theta_c$ is the correction angle, and $Az_F$ is final calculated azimuth from the survey, measured with respect to true north. $\Delta Az$ is the difference between $Az_{SC}$ and $Az_F$. All azimuths are rounded to the nearest degree with an error of 5°. The northerly row at site 44-6-0019 had been destroyed by bulldozing as noted in the site card.

| Site Card | $Az_{SC}$ | $D_{mag}$ | $Az_S$ | m | $\chi^2$ | $\theta_c$ | $Az_F$ | $\Delta Az$ |
| --- | --- | --- | --- | --- | --- | --- | --- | --- |
| | (°) | (°) | (°) | (°) | | (°) | (°) | (°) |
| 20-2-0005 | 133 | 11.2 | 108 | +0.0051 | 0.07 | +0.3 | 108 | 25 |
| 27-2-0004 | 90 | 10.5 | 89 | −0.0313 | 0.24 | −1.8 | 87 | 3 |
| 28-2-0005 | 181 | 11.0 | 165 | −0.0079 | 0.24 | −0.5 | 165 | 16 |
| | 183 | 11.0 | 175 | −0.0182 | 0.12 | −1.0 | 174 | 9 |
| 44-6-0019 | 0 | 12.0 | | | | | | |
| | 90 | 12.0 | 63 | −0.0064 | 0.00 | −0.4 | 63 | 27 |
| 51-1-0053 | 180 | 11.8 | 143 | −0.0018 | 0.05 | −0.1 | 143 | 37 |

We now determine whether or not the orientations of the entire dataset can be accounted for by chance. To account for the uncertainty in whether or not the azimuths were given with respect to true or magnetic north, we present the survey data with the site card data in 10° bins and smooth it using a 3-bin boxcar convolution (to account for $\theta \approx 20° \pm 12°$), given as:

$$\text{bin n} = \left[ \frac{\text{bin (n − 1)} + \text{bin (n)} + \text{bin (n + 1)}}{3} \right]$$

Where *n* is each given bin in the histogram. The smoothed data is given in Figure 4. For the unsmoothed (original) data, $10^5$ Monte Carlo simulations were performed, of which 650 attained a peak of 10 in the 0° or 90° bins. This gives a chance probability of 0.7%. For the smoothed data, the same number of simulations was performed, attaining a peak of 4.3 in the 0° bin 115 times, giving a probability of 0.1%. In either case, it is clear that these alignments are not the product of chance; there is a clear preference for these azimuth ranges.

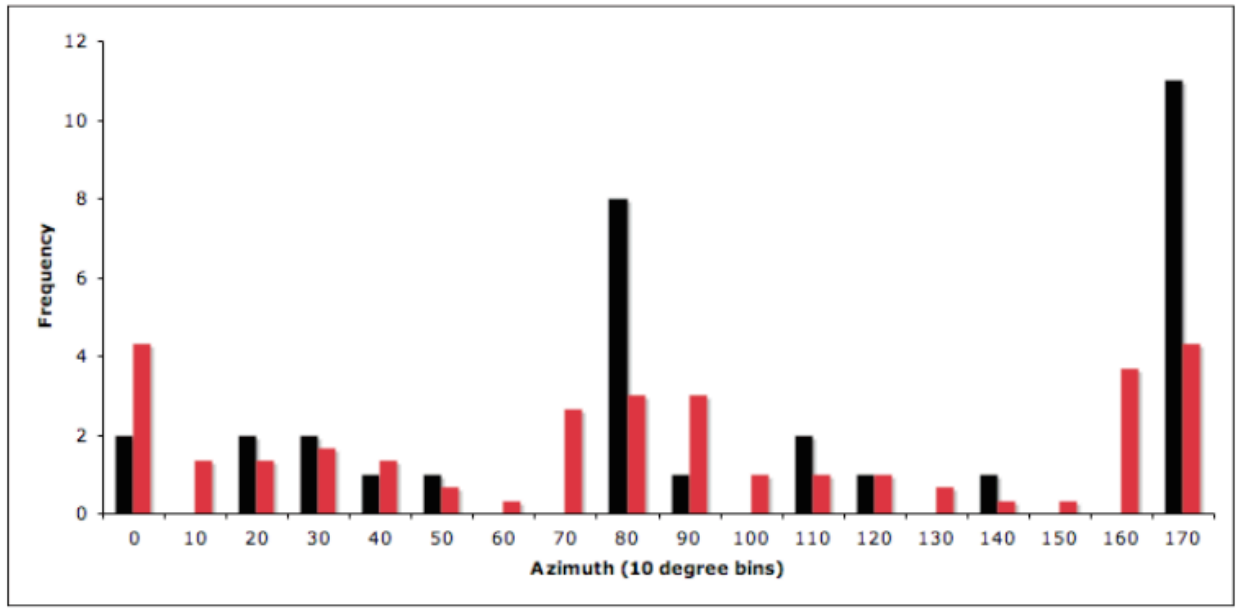

**Figure 4:** Histogram showing the azimuths (black) and the smoothed data (red) in 10º bins. This represents the entire data set corrected for Dmag, assuming all site cards gave magnetic north, as indicated by the analysis. Data is smoothed to account for the ~20º uncertainty from θ.

## Discussion

From these results, it is apparent that there are two preferred orientations for linear stone arrangements: a lesser preference towards east/west and a greater preference towards north/south. To explain the preferred cardinal orientations, we consider three possibilities:

1. The stone rows were constructed by Europeans with a preferred orientation to cardinal points;
2. The stone rows were systematically drawn or described by the surveyors as having preferred cardinal orientations, despite the stone rows not actually having this preference; and,
3. The stone rows were constructed by Aboriginal people with a preferred orientation to cardinal points.

Possibility 1 depends on whether or not the stone arrangements are Aboriginal in origin. The first criterion in selecting stone rows for this study is that they were constructed by Aboriginal people. Some of the site cards identified arrangements that were European in origin (e.g. survey lockspits). These site cards were rejected during the selection process and all of the sites that passed this criterion are argued to be Aboriginal in origin based on their morphology, proximity to other Aboriginal artefacts (such as scarred trees and rock art), or their identification as such by members of the local Aboriginal community. Thus, we reject the position that these arrangements are European in origin.

Possibility 2 depends on the accuracy of information recorded on the site cards. The surveyors state in several of the site cards that their drawings are not to scale and were only rough sketches or descriptions. Since $\theta = 19.5°\pm11.5°$, the mean magnetic declination ($\mathbf{D_{mag}}$) is within $1$-$\sigma$ of $\mathbf{D_{mag}}$ at the sites surveyed (11.3°), suggesting the surveyors are reasonably accurate in recording their findings, except that they generally cited magnetic north instead of true north.

Possibility 3 is that the arrangements are Aboriginal in origin and that Aboriginal people oriented stone rows to cardinal directions, both of which are supported by the site card information and the data analysis. However, we are puzzled as to why these arrangements seem to be oriented to magnetic north. There is currently no evidence that Aboriginal people used any form of magnetic compass. Since we have determined that the arrangements are not European in origin, it indicates that the bias may lie with the surveyors. It is probable that the surveyors used a compass and took a general orientation, but did not place emphasis on high accuracy (which is why we say 'reasonably accurate'). We do not know what percentage of the site cards gave magnetic or true north without revisiting each site. The smoothed data accounts for this uncertainty and shows us that preferred cardinal orientations are within the error limit.

This paper is a preliminary test to see if linear stone arrangements have a preferred orientation to the cardinal points. Our analysis suggests that they do, but we were only able to survey a small subset of sites and are not certain of the reliability of the remaining site cards. Future work will involve locating and surveying as many arrangements as reasonably possible to better constrain our estimates. The geographical range of the sites surveyed in this paper is significant, ranging from the northern tablelands near Armidale to the southern slopes near Canberra and from the coast to central NSW. Most of the stone arrangements belong to different Aboriginal language groups. To overcome uncertainties, it would be ideal to survey a number of stone arrangements within a particular language group to see if these orientations remain statistically significant.

Future work should also focus on understanding how the orientations were estimated and what the purpose was in orienting stone arrangements to these points. We currently do not know the reason these arrangements were oriented to the cardinal points, but we can suggest ways these points were estimated. It may be that east/west orientations are related to the rising and setting position of the sun in the sky as noted by some Aboriginal cultures, such as the Wangkumara and Yirandhali. Other stone arrangements in Australia have been found with precise east/west orientations, such as Wurdi Youang (Hamacher and Norris 2011). North and south are 90° to east and west, and the concept of right angle orientations is found in some Aboriginal cultures (e.g. the Warlpiri and Guugu Yimithirr). South can be roughly determined at night by locating the south celestial pole and looking in the direction of the horizon directly below this point. A number of techniques can be used to accomplish this, such as noting the midpoint of a circumpolar star. An Aboriginal man and educator from Port Lincoln, South Australia, said that Venus was a good indicator of west, as it shone brightly at dusk and is near the sun (Pring 2002:9). Another Aboriginal man from South Australia said he could tell direction at night based on the position of the Milky Way—it would stretch from east to west during the winter, and from north to south during the summer (Pring 2002:9). These techniques are approximations of finding the cardinal directions; they are only accurate to within several degrees. For example, the position of the rising or setting sun on the horizon will vary by nearly 30° either side of due east or west. These extreme rising/setting positions occur at the summer and winter solstice. The only time the sun rises exactly at due east or sets due west is at the vernal or autumnal equinox.

We cannot currently say with any certainty why these structures were oriented to the cardinal directions or how those directions were determined, but examples across Australia give us clues. The concept of cardinal directions is found in the cultures of several Aboriginal groups and techniques for finding the cardinal points based on the positions of astronomical bodies are known, providing plausible explanations on which to base future work.

**Acknowledgements**

We would like to acknowledge the traditional custodians and landowners of the sites we visited, the NSW Department of Environment & Heritage, Craig O'Neill, the Department of Indigenous Studies at Macquarie University for funding the project, and the Department of Earth and Planetary Science at Macquarie University for loaning us the survey equipment.